\documentclass[aps,prd,twocolumn,nofootinbib,floatfix,preprintnumbers]{revtex4}
\usepackage{subfigure}
\usepackage{graphicx}
\usepackage{dcolumn}
\usepackage{epsfig}
\usepackage{amsmath}
\usepackage{amsfonts}
\usepackage{amssymb}
\usepackage{color}
\usepackage{hyperref}
\usepackage{bm}

\newcommand{\be}{\begin{equation}}
\newcommand{\ee}{\end{equation}}
\newcommand{\bea}{\begin{eqnarray}}
\newcommand{\eea}{\end{eqnarray}}

\begin{document}

\title{Basis function expansions of the two-point correlation function\\ and reconstruction of the cosmological distance scale}

\author{Fatima Abd Alrahman${}^{1,2,3}$}
\author{Farnik Nikakhtar${}^4$}
\author{Marcello Musso${}^{1,5}$}
\author{Aseem Paranjape${}^6$}
\author{Ravi K.~Sheth${}^{7}$}
\affiliation{${}^1$ICTP East Africa Institute for Fundamental Research, University of Rwanda, Kigali -- Rwanda}
\affiliation{${}^2$Department of Physics, University of Houston, 3507 Cullen Blvd Rm 617, Houston, TX 77204}
\affiliation{${}^3$Department of Physics, University of Khartoum, Al-Gama'a Avenue, Khartoum, Sudan}
\affiliation{${}^4$Department of Physics, Yale University, New Haven, CT 06511, USA}
\affiliation{${}^5$Departamento de F{\'i}sica Fundamental \& IUFFyM, Universidad de Salamanca, 37008 Salamanca, Spain}
\affiliation{${}^6$Inter-University Centre for Astronomy \& Astrophysics, Ganeshkhind, Post Bag 4, Pune 411007 -- India}
\affiliation{${}^7$Center for Particle Cosmology, University of Pennsylvania, 209 S. 33rd St., Philadelphia, PA 19104 -- USA}

\date{\today}

\begin{abstract}
  The shape of the correlation function of galaxy pairs can be used as a cosmological distance indicator.  Most estimators of this shape require that one bin the pair counts, and considerable effort is expended in finding the optimal bin size.  Cosmological constraints are then derived, in a separate step, by fitting to the binned pair counts.  By projecting the pair counts on a suitable basis set, it is possible to estimate the pair correlation function without binning.  Recent work suggests that, in the context of estimating the cosmological distance scale from galaxy clustering data, polynomials and half-integer generalized Laguerre functions are particularly useful bases.  We show how to use these bases to measure the correlation function and estimate the reconstructed distance scale in a single step.  Our analysis includes the effects of shot-noise and cosmic variance, and highlights the benefits of projecting pair counts onto orthogonal basis functions.  
\end{abstract}


\maketitle

\newcommand{\ste}[1]{\textcolor{red}{\textbf{\small[Ste: #1]}}}


\section{Introduction}\label{sec:intro}

The shape of the correlation function of galaxy pairs encodes a wealth of information about galaxy formation and cosmology \cite{cs02, fullshapePk}.  For instance, the peak and dip on $\sim 150$Mpc (comoving) scales can be used as a cosmological distance indicator \cite{esw2007,PaperI,modelAgnosticAP}.  Current and planned datasets have the potential to provide percent-level precision constraints \cite{alamDR12,sanchezDR12,LPdesi}.  For these constraints to be accurate, one must ensure that the estimator of the correlation function signal itself is unbiased.  To date, the vast majority of estimators work with binned pair counts \cite{ls93}.  However, determining the optimal bin size is tedious \cite{boss2014,PRDmocks}.  Recently, `least squares' estimators which do not require binning, have been proposed \cite{lsqXi18,nobinXi21}.  These expand the correlation function in a set of basis functions, but do not address the question of what basis functions to use.  Refs.\cite{LPlaguerre,HODlaguerre} showed that generalized half-integer Laguerre functions are a particularly interesting choice for studies of the cosmic distance scale.  This is because such studies typically have two steps:  first, the binned correlation function is measured from the data, and then a model is fit to the binned measurements, from which constraints are finally derived.  In principle, working with generalized half-integer Laguerre functions should allow one to merge the two steps into one.  The main goal of the present study is to explore this synergy.  

Section~\ref{sec:why} describes our formulation of the no-bins, least-squares method.  We treat the ideal case of noiseless data first, before showing how to account for the correlated errors which arise from discreteness and finite survey size.  Section~\ref{sec:what} shows our results and a final section summarizes.  Our approach is particularly transparent when the basis functions are orthogonal; this is the subject of Appendix~\ref{sec:gso}.  Appendix~\ref{sec:smearing} provides a detailed discussion of the choice of basis functions when the goal is to both estimate and deconvolve the pair correlation function.  

\section{The Method}\label{sec:why}

\subsection{Idealized case}
We start by defining the fraction of distinct pairs with separation $r_{ij}$ that lie in a tophat bin of width $\Delta r$ centered on $r_k$:  
\begin{equation}
  DD_k = \frac{2}{N(N-1)}\sum_{i=1}^N\sum_{j>i}^N I_k(|r_{ij}|) ,
  \label{eq:DDbinned}
\end{equation}
where $I_k(|r_{ij}|)=1$ if $r_{ij}$ lies within $\pm\Delta r/2$ of some $r_k$ (i.e. if $r_{ij}$ lies in bin $k$), and is zero otherwise.  We can generalize slightly to 
\begin{equation}
  DD_k \equiv \frac{2}{N(N-1)}\sum_{i=1}^N\sum_{j>i}^N f_k(r_{ij}) ,
  \label{eq:DDunbinned}
\end{equation}
which is a sum over all $N(N-1)/2$ distinct pairs in the data, with each pair given a weight, $f_k(r_{ij})$, that depends on the pair separation $r_{ij} \equiv |{\bm r}_i - {\bm r}_j|$.  As the sum is normalized by the total number of distinct pairs, $DD_k$ is the fraction which $f_k$ contributes to the total pair count.  
In contrast to the `binned' estimator of equation~(\ref{eq:DDbinned}), `unbinned' estimators really just relax the assumption about the (rectangular tophat) bin shape, allowing it to be arbitrary.  

This raises the question of what bin shapes are `optimal', or `physically motivated'.  We will address this shortly.  However, whatever the shape, the theorist would write the sum above as 
\begin{equation}
 d_k \equiv \frac{4\pi}{V_{\rm tot}} \int dr\,r^2 \, \Big[1 + \xi(r)\Big]\,f_k(r),
 \label{eq:dk}
\end{equation}
where $\xi(r)$ is the pair correlation function and $V_{\rm tot}$ is the total volume in which the pairs were counted.  If we now assume that, over the range $r_{\rm min}$ to $r_{\rm max}$, the pair correlation function $\xi(r)$ can be well approximated using a linear combination of $n+1$ basis functions $f_j(r)$: 
\begin{equation}
 \xi(r) = \sum_{j=0}^n a_j\,f_j(r),
 \label{eq:xiBasis}
\end{equation}
then 
\begin{equation}
 d_k = v_k + a_j F_{jk},
\end{equation}
where repeated indices must be summed over, and we have defined 
\begin{equation}
 v_k \equiv \frac{4\pi}{V_{\rm tot}} \int_{r_{\rm min}}^{r_{\rm max}} dr\,r^2\,f_k(r)
 \label{eq:vk}
\end{equation}
and
\begin{equation}
  F_{jk} \equiv \frac{4\pi}{V_{\rm tot}}
               \int_{r_{\rm min}}^{r_{\rm max}} dr\,r^2\,f_j(r)\,f_k(r).
 \label{eq:Fjk}
\end{equation}
(The usual binned estimator has $F_{jk} = \delta_{jk}\,v_k$.) 
If the model were able to provide a perfect descripton of $\xi$, and the dataset was so large (i.e. both $N$ and $V_{\rm tot}$ are large) that the sum over pairs is affected by neither discreteness nor finite volume effects, then one expects to have
\begin{equation}
  DD_k = d_k = v_k + a_j F_{jk}.
\end{equation}
Provided that no two basis functions have the same shape, this expression can be inverted to yield
\begin{equation}
  (DD_k - v_k)\,F^{-1}_{kl} = a_jF_{jk}F^{-1}_{kl} = a_l.
  \label{eq:simple}
\end{equation}
The left hand side of this expression gives the coefficients $a_j$ in terms of $DD_k$ which are measured, and $v_k$ and $F_{jk}$ which depend on the choice of basis functions, so are also known.  Once determined, the $a_j$ can be inserted in equation~(\ref{eq:xiBasis}) to estimate $\xi$.  

\subsection{Shot-noise, cosmic variance, and error bands}
In practice, finite $N$ brings counting errors (which are typically assumed to be Poisson-distributed and are often called `shot-noise'), and finite $V_{\rm tot}$ brings `cosmic variance' errors.  As a result, in practice,
\begin{equation}
  DD_k - d_k = e_k \ne 0,
\end{equation}
where $e_k$ represents the `error'.
Typically, the error has $\langle e_k\rangle=0$, where the average is taken over many different realizations of the data.  However, the covariance matrix of second moments $E_{ij}=\langle e_ie_j\rangle$ is typically non-zero.  (In some cases $E_{ij}$ may be diagonal, but we will not assume this in what follows.)

It is then natural to define
\begin{equation}
  \chi^2 \equiv
  \sum_{ij} e_i E_{ij}^{-1} e_j = 
  \sum_{ij} (DD_i - d_i) E_{ij}^{-1} (DD_j - d_j),
  \label{eq:chi2}
\end{equation}
so that the $a_j$ are now determined by minimizing $\chi^2$.  In matrix notation,
\begin{align}
  \frac{\partial\chi^2}{\partial {\bm a}} &=
  \frac{\partial ({\bm{DD}} - {\bm v} - {\bm a}^T{\bm F})^T {\bm E}^{-1} ({\bm{DD}} - {\bm v} - {\bm a}^T{\bm F})}{\partial {\bm a}} \nonumber\\
  &= -2 {\bm F}{\bm E}^{-1} ({\bm{DD}} - {\bm v} - {\bm a}^T{\bm F}).
\end{align}
This system of equations is minimized when 
\begin{equation}
  {\bm F}{\bm E}^{-1} ({\bm{DD}} - {\bm v})
  =  {\bm F}{\bm E}^{-1} {\bm a}^T{\bm F}
  =  {\bm F}{\bm E}^{-1} {\bm F}^T {\bm a}
\end{equation}
which, in turn, yields 
\begin{align}
  {\bm a} &= \Big({\bm F}{\bm E}^{-1} {\bm F}^T\Big)^{-1} {\bm F}{\bm E}^{-1}
              ({\bm{DD}} - {\bm v})\nonumber\\
          &= \Big({\bm F}^T\Big)^{-1} \Big({\bm F}{\bm E}^{-1}\Big)^{-1} {\bm F}{\bm E}^{-1}
              ({\bm{DD}} - {\bm v}) \nonumber\\
    &= \Big({\bm F}^T\Big)^{-1} ({\bm{DD}} - {\bm v}),
  \label{eq:full}
\end{align}
which is just equation~(\ref{eq:simple}).

Equation~(\ref{eq:full}) makes obvious that minimizing $\chi^2$ to determine ${\bm a}$ is really just a generalized least squares estimation problem.
This correspondence means that the covariance matrix of the fitted coefficients, Cov$(a_i,a_j) \equiv A_{ij}$, is given by 
\begin{equation}
  {\bm A} = \Big({\bm F}{\bm E}^{-1} {\bm F}^T\Big)^{-1}.
  \label{eq:covA}
\end{equation}
The covariance matrix ${\bm A}$ provides confidence bands around the $\xi$ which one gets from inserting the best-fit ${\bm a}$ in equation~(\ref{eq:full}).  Standard propagation of errors means that
\begin{align}
  \Big\langle\xi(r)\,\xi(s)\Big\rangle &= {\bm f(r)}^T {\bm A}\,{\bm f}(s) \nonumber\\
   &={\bm f(r)}^T ({\bm F}^T)^{-1} {\bm E}\,{\bm F}^{-1}\,{\bm f}(s).
  \label{eq:covXiNL}
\end{align}
Of course, these values and their uncertainties are only interesting if $\chi^2_{\rm min}$ of equation~(\ref{eq:chi2}) is close to zero 
(if not, then the chosen basis does not provide a good fit to the measurements, so the values of the best-fit $a_j$ are uninteresting).

All that remains is to specify ${\bm E}$.  Its elements are
\begin{align}
  E_{ij} = \frac{2}{V_{\rm tot}} \int_0^\infty \frac{dk\,k^2}{2\pi^2}\,
            \Big[P(k) + \frac{1}{\bar{n}}\Big]^2\,w_i(k) \, w_j(k),
  \label{eq:Eij}
\end{align}
where 
\begin{equation}
  w_j(k) = \frac{4\pi}{V_{\rm tot}}
           \int_{\rm r_{min}}^{r_{\rm max}} dr\, r^2\,f_j(r)\,j_0(kr),
\end{equation}
$P(k)\approx b^2 P_{\rm Lin}(k)\,e^{-k^2\Sigma^2}$ with $\Sigma^2\approx\int dk\,P_{\rm Lin}(k)/(3\pi^2)$ \cite[on large BAO scales; e.g.][]{rpt}, and $\bar{n} \equiv N/V_{\rm tot}$.  

To gain intuition, recall that the usual binned estimator has diagonal $F_{jj}=v_j$.  If $P(k)\ll 1/\bar{n}$ then we show in Appendix~\ref{sec:gso} that ${\bm E}\to E_{jj} = (2/N^2)\,v_j$ is also diagonal, so 
$\langle\xi(r)\,\xi(s)\rangle\to \langle\xi(r_i)\,\xi(r_j)\rangle\to E_{jj}/v_j^2 = (2/N^2)/v_j$. The presence of $v_j$ in the denominator is responsible for the well-known divergence which appears in the limit of narrow bins.  Our `unbinned' estimator avoids this divergence.  

In this context, it is worth noting that there are three special but distinct ways of organizing a given set of basis functions:  diagonalize $\bm{A}$, $\bm{F}$ or $\bm{E}$.  Although diagonalizing $\bm{A}$ is a popular choice that is determined by the (expected) properties of the data as well as of the basis function shapes, diagonalizing $\bm{F}$ is attractive because then 
$\langle\xi(r)\xi(s)\rangle \to (f_i(r)/F_{ii})\,E_{ij}\, (f_j(r)/F_{jj})$.  We discuss other benefits of this choice in Appendix~\ref{sec:gso}. 

This is a good place to highlight the following conceptual point.  Traditional analyses would fit a parametrized model to the binned estimate of $\xi$.  This binned estimate would typically {\em not} require prior knowledge of the dataset -- the bin-to-bin covariances -- but the subsequent fitting step would.  There is a sense in which this is attractive, since it appears to make the measurement step `independent' of the step in which a model is fit to the measurements.  In our method, the two appear to have been combined into a single step, so it is interesting to ask how the covariance enters.  
We used $\bm{E}$ to denote this covariance (equation~\ref{eq:Eij}) and the appearance of $\bm{E}$ in the first line of  equation~(\ref{eq:full}) makes it appear that the estimation step requires prior knowledge of the `theory'.  However, the final line of equation~(\ref{eq:full}) shows that the $\bm{E}$-dependence drops out.  Moreover, this final line is the {\em same} as equation~(\ref{eq:simple}) in which $\bm{E}$ was not used at all.  Thus, we have shown that the mean of the estimated shape of $\xi$ does {\em not} depend on $\bm{E}$.  In this respect, one could think of the projection of pair counts onto basis functions as a form of data compression, rather than as model `fitting'.  In contrast, the {\em uncertainties} on the estimated coefficients (equation~\ref{eq:covA}), and the estimated error bands which indicate the range of allowed shapes (equation~\ref{eq:covXiNL}), do depend explicitly on $\bm{E}$.  This is analogous to the traditional `bin+fit' case in which the fitting process (the chi-square that is minimized when fitting) and derived error bands depend on prior knowledge of the process (e.g., the expected cosmic variance).  Of course, one could try to fit these (or other) basis functions to the amplitudes of the projected counts:  this would be the analog of fitting, e.g., polynomials to the traditional binned estimator, and {\em would} require use of the covariance matrix.  


\subsection{Basis functions for pre- and post-reconstruction distance scale estimates} 
The analysis above is general -- we have still not specified the basis functions $f_j$.  Ref.\cite{LPlaguerre} shows that, in the context of distance scale measurements, half-integer generalized Laguerre functions -- which they denote $\mu_j(r)$ -- are interesting because if
\begin{equation}
  \xi_{\rm NL}(r) = \sum_j a_j\,\mu_j(x)  \quad {\rm then}\quad
  \xi_{\rm Lag}(r) \equiv \sum_j a_j\,x^j 
  \label{eq:xiLag}
\end{equation}
with $\xi_{\rm Lag}\approx \xi_{\rm Lin}$ on BAO scales. 
Here $x = r/\Sigma$ with $\Sigma$ a `smearing' scale associated with evolution from the initial conditions to the time of observation.  (In what follows, we assume that this scale is known, but see \cite{modelAgnosticAP} for more discussion.)
See \cite{bayesLP} for how to determine the appropriate number of terms in the sum. 

We show in Appendix~\ref{sec:smearing} that this motivates the setting of $f_j=\mu_j$.  Then 
 $\xi_{\rm NL}(r) = \bm{a}^{\rm T}\bm{f}(r)$, 
and (the square-root of) $\bm{f}^{\rm T}(r)\bm{A}\bm{f}(r)$ gives the error band around the best fit.  Moreover, if we use $p_j$ to denote the polynomial $r^j$, then one can reconstruct $\xi_{\rm Lag}$ from this fit by setting 
 $\xi_{\rm rec} = \bm{a}^{\rm T}\bm{p}$ 
(i.e., the same coefficients $a_j$ now multiply $p_j$ rather than $\mu_j$), with error bands given by (the square-root of) $\bm{p}^{\rm T}(r)\bm{A}\bm{p}(r)$.  Appendix~\ref{sec:smearing} shows how to proceed if one wishes to use $p_j$ rather than $\mu_j$ as basis functions for the no-bins estimate.  

In addition, the linear point in the correlation function, 
\begin{equation}
  r_{\rm LP} \equiv \frac{r_{\rm peak} + r_{\rm dip}}{2},
  \label{eq:rLP}
\end{equation}
where $r_{\rm peak}$ and $r_{\rm dip}$ are those scales where $d\xi/dr=0$, is a standardizable rod for cosmological distance scale analyses \cite{PaperI,LPlaguerre}.  Standard propagation of errors yields its uncertainty:
\begin{equation}
   \sigma_{\rm LP}^2 = \sum_{i,j} \frac{\partial r_{\rm LP}}{\partial a_i}
    \, A_{ij}\, \frac{\partial r_{\rm LP}}{\partial a_j}
  \label{eq:sigLP}
\end{equation}
\cite{LPnus,LPlaguerre}, where $r_{\rm LP}$ is that nonlinear combination of the $a_k$ and $f_k(x)=\mu_k(x)$ functions which comes from requiring $\xi_{\rm NL}'=0$, and the covariance matrix ${\bm A}$ was defined in equation~(\ref{eq:covA}).

Similarly, $r_{\rm LP-rec}$ can be measured from the peak and dip scales in the Laguerre reconstructed correlation function $\xi_{\rm rec}$ (i.e. where $\xi_{\rm rec}'=0$, where $\xi_{\rm rec}$ uses the same $a_k$ but sets $f_k=p_k$), and the uncertainty on its value is given by the expression above, so that the covariance $\bm{A}$ again matters, but $\partial r_{\rm LP-rec}/\partial a_j$ is different (because it is determined for the peak and dip scales of a different function).   

This shows explicitly that different combinations of the quantities determined from measuring $\xi_{\rm NL}$ yield $\xi_{\rm rec}$ and associated error bands.  Thus, by measuring $\xi_{\rm NL}$ using Laguerre-functions, one effectively combines the measurement and reconstruction steps associated with BAO analyses.  

\section{Results}\label{sec:what}
We validate our methodology using measurements of the evolved correlation function of massive dark matter halos identified in 20 different $z=0.5$ outputs of the Abacus simulation suite of \cite{abacus}.  Specifically, we use the HM halo samples which were studied in \cite{LPlaguerre}.  Each box has volume $V_{\rm tot}=(1.1h^{-1}$Gpc$)^3$, with halo number density $\bar{n}=8.6\times 10^{-5}(h/{\rm Mpc})^3$ and bias factor $b=2.6$.

\begin{figure}
    \centering
    \includegraphics[width=\linewidth]{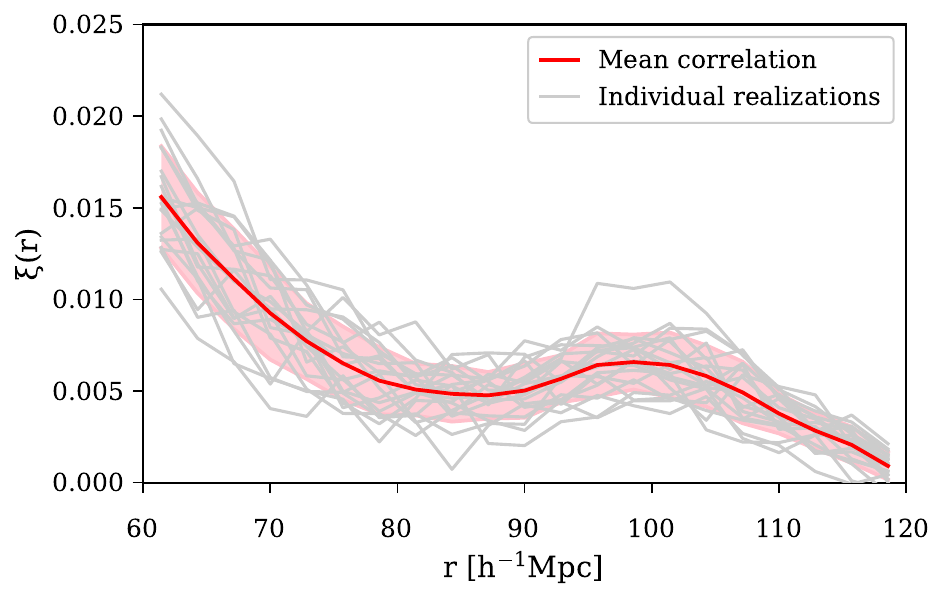}
    \caption{Binned correlation functions measured in 20 realizations of the Abacus simulation suite (each box is $1.1h^{-1}$Gpc on a side); the bold curve shows the mean of these 20 realizations, and the pink band shows the rms scatter around this mean.  In each realization, the pairs were counted in 20 tophat, non-overlapping bins, each $3h^{-1}$Mpc wide, running from $60-120h^{-1}$Mpc.}
    \label{fig:xibinned}
\end{figure}

Figure~\ref{fig:xibinned} shows the same binned correlation functions $\xi_{\rm NL}$, measured in non-overlapping bins running from $60-120h^{-1}$Mpc, each $3h^{-1}$Mpc wide that were used in Ref.~\cite{LPlaguerre}.  The thick solid curve shows the average of these 20 realizations.  Each realization corresponds to a comoving volume of $(1.1h^{-1}$Gpc)$^3$ so the thick solid curve represents a measurement in a volume of $\sim 25h^{-3}$Gpc$^3$, which is comparable to a number of ongoing surveys.

\begin{figure}
    \centering
    \includegraphics[width=\linewidth]{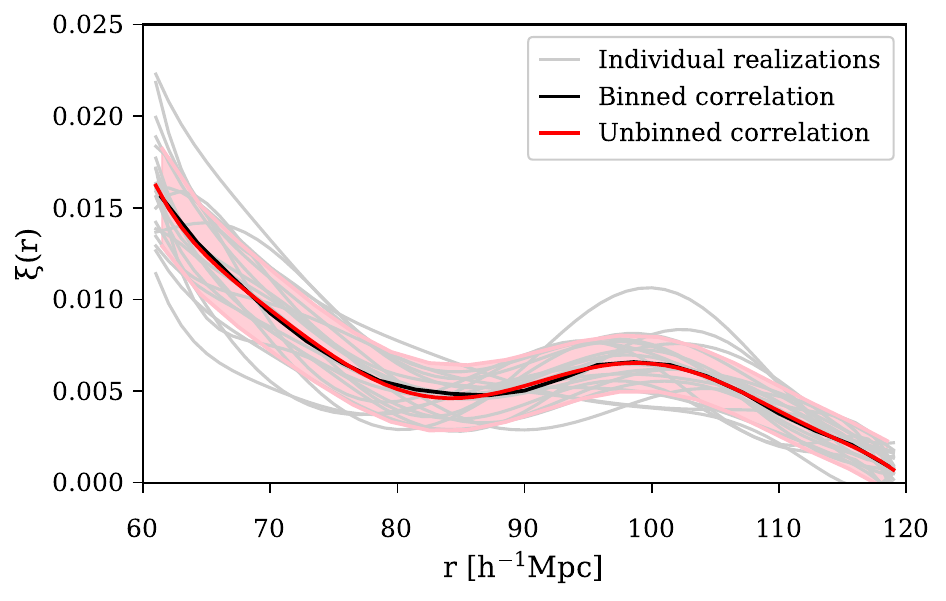}
    \caption{Unbinned correlation functions $\xi_{\rm NL}$ measured in the same 20 realizations as the previous figure; in this case, the pairs were `counted' in 10 `bins' whose shapes are determined by the 9th lowest order half-integer modified Laguerre polynomials defined in \cite{LPlaguerre}.  The smoothness of the basis functions translates into smoother estimates of $\xi_{\rm NL}$ in each realization.  Thick dashed curve shows the mean of these 20 curves, error bands show the uncertainty on a single realization (the uncertainty on the mean would be $\sqrt{20}\times$ smaller), and thick solid curve (same as Figure~\ref{fig:xibinned}) shows the mean of the binned estimates of $\xi_{\rm NL}$. }
    \label{fig:xinobins}
\end{figure}

Each of these curves, and their average, should be compared with corresponding curves shown in Figure~\ref{fig:xinobins}.  These were made by projecting the counts onto the Laguerre-basis discussed in Ref.~\cite{LPlaguerre}.  (This basis is specified by a single fiducial scaling parameter ${\cal R}$ which they, and so we, set equal to 4.6$h^{-1}$Mpc.)  We used 10 basis functions, so the total number of parameters required to describe $\xi_{\rm NL}$ is half that associated with the tophat bins.  We determined the coefficients $a_k$ using equation~(\ref{eq:simple}).
The basis functions are smooth functions of scale, so each curve in Figure~\ref{fig:xinobins} is smoother than its counterpart in Figure~\ref{fig:xibinned}.  Nevertheless, the mean shape, shown by a thick dashed curve, is very similar to the mean of the binned estimator, the thick solid curve of Figure~\ref{fig:xibinned}.  (Note also that the curves in Figure~\ref{fig:xibinned} are shown as functions of bin center, whereas the curves in Figure~\ref{fig:xinobins} can be shown as functions of any scale.)  The bands around the mean curve show the estimated uncertainty on a single realization  (equation~\ref{eq:covXiNL} with $r=s$, and the values of $\bar{n}$, $b$ and $V_{\rm tot}$ given earlier; the error on the mean curve would be smaller by a factor of $\sqrt{20}$).  If we work with centered Laguerres of Ref.~\cite{LPlaguerre}, then the error bar is dominated by the contribution from the first few lowest order terms (the elements in the upper left corner of the matrix ${\bm A}$ are much larger than the others).  We have also projected onto the functions defined by orthogonalizing the centered Laguerres (see discussion in Appendix~\ref{sec:gso}) and verified that the fits and associated error bands are indistinguishable from those shown in Figure~\ref{fig:xinobins}. 

\begin{figure}
    \centering
    \includegraphics[width=\linewidth]{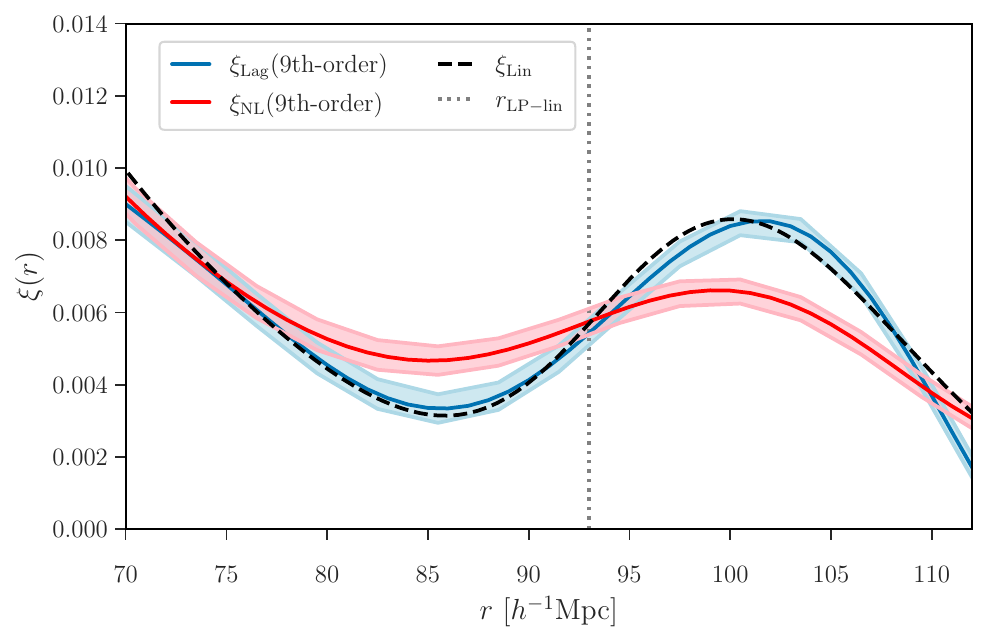}
    \caption{Comparison of the linear theory shape (dashed) with the 9th order half-integer Laguerre function descriptions of $\xi_{\rm NL}$ (red) and the associated reconstructions $\xi_{\rm Lag}$ (blue).  Error bands estimated as described in the text assume a total volume of $20\times (1.1h^{-1}$Gpc)$^3$.}
    \label{fig:xirecon}
\end{figure}

Having demonstrated that the estimated $\xi_{\rm NL}$ are reasonable, both for the individual realizations and for their mean, we will now consider the mean, which represents a measurement of $\xi_{\rm NL}$ in a comoving volume of $\sim 25h^{-3}$Gpc$^3$.  We are particularly interested in the reconstructed shape $\xi_{\rm Lag}$ whose shape is related to that of $\xi_{\rm NL}$ by equation~(\ref{eq:xiLag}).  
Figure~\ref{fig:xirecon} shows an estimate of $\xi_{\rm NL}$ which includes terms upto 9th order in $\mu_k$.  It is very different from $\xi_{\rm Lin}$ -- the shape from which unbiased constraints on the cosmological distance scale are most easy to derive.  The other curve shows the associated reconstructed shape $\xi_{\rm Lag}$ (error bands computed as described in the text between equations~\ref{eq:xiLag} and~\ref{eq:rLP}); it is much closer to $\xi_{\rm Lin}$ (dashed curve).  (The agreement is particularly good because we have fixed the smearing scale $\Sigma$ to its true value.)  To quantify the difference between $\xi_\text{NL}$ and $\xi_\text{Lag}$, we have estimated the associated $r_\text{LP}$ values.  They agree with the HM values shown in Table~1 of \cite{LPlaguerre}.  In particular, $r_\text{LP}$ for $\xi_\text{Lag}$ is very close to that associated with $\xi_{\rm Lin}$, which is shown as a vertical dashed line in Figure~\ref{fig:xirecon}.  

\begin{figure}
    \centering
    \includegraphics[width=\linewidth]{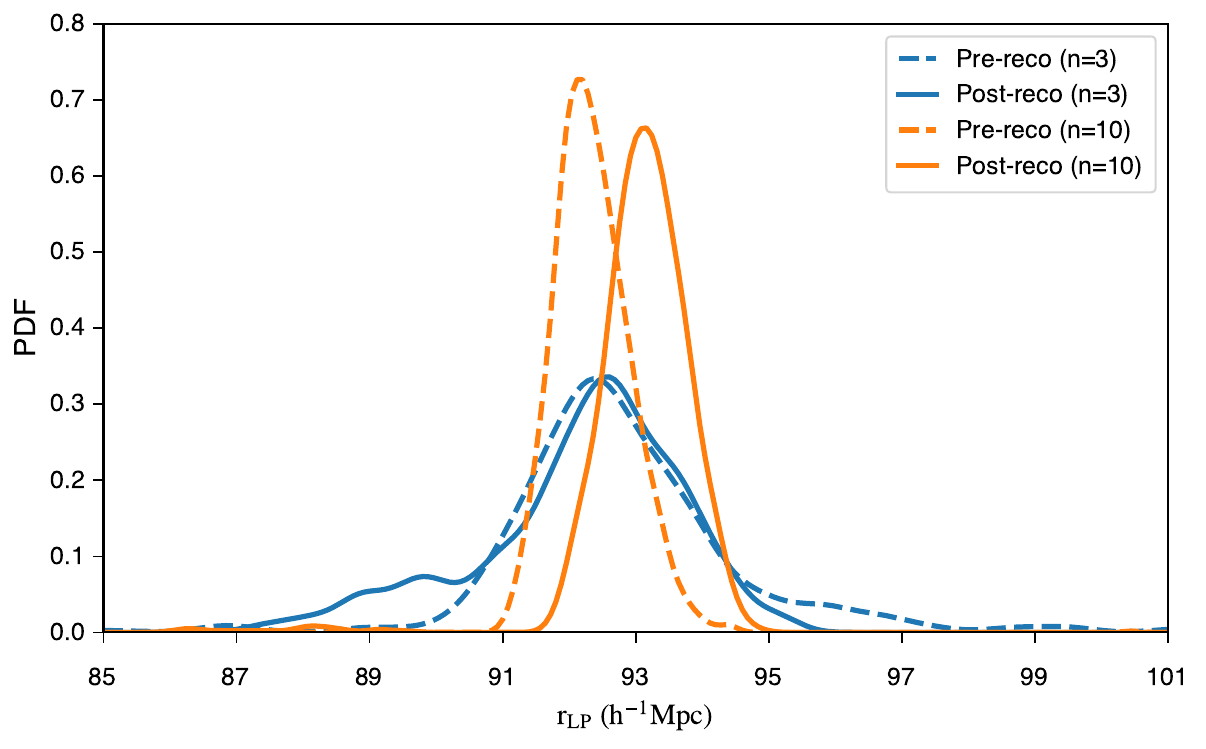}
    \caption{Distribution of $r_\text{LP}$ values in 1000 realizations of an effective volume of $4h^{-3}$Gpc$^3$ and $13h^{-3}$Gpc$^3$ from projecting onto Laguerre basis functions (dashed) and associated `reconstructed' simple polynomials (solid).  Larger effective volumes result in narrower distributions, for which the reconstructed value becomes centered on the linear theory value of $93h^{-1}$Mpc.}
    \label{fig:rLPvEff}
\end{figure}

To illustrate how our results scale with effective volume, Figure~\ref{fig:rLPvEff} shows the result of approximating a larger effective volume by combining estimates from smaller volumes.  Here, we randomly choose $n$ simulation boxes (without replacement), average their $a_k$ values, and estimate $r_\text{LP}$ in the associated pre- and post-reconstructed correlation function shapes (i.e. $\xi_\text{NL}$ and $\xi_\text{Lag}$).  We repeat this $10^3$ times to get a distribution of values.  We did this for $n=3$ and $n=10$. (Since we have only 20 simulation boxes, for $n=20$ there is only one realization, and it is shown in Figure~\ref{fig:xirecon}.)  Figure~\ref{fig:rLPvEff} shows that, for the larger effective volumes the distributions are tighter and the `reconstructed' values (i.e., those for $\xi_\text{Lag}$) are more closely centered on the linear theory value.  

In all cases, the $r_{\rm LP}$ from binned and unbinned estimates agree within their uncertainties, and similarly for $r_{\rm LP-rec}$ (i.e., fitting Laguerres to a binned $\xi_{\rm NL}$ measurement, and then using the fitted coefficients to estimate a reconstructed shape from which to estimate $r_{\rm LP-rec}$, versus determining the Laguerre coefficients directly from the unbinned pair counts). These uncertainties are smaller than if we had we estimated from (differentiating) the binned counts directly (i.e. without fitting a smooth functional form to the binned counts).

Since the $\xi_{\rm Lag}$ shape uses the same parameters $a_k$ that were determined by the estimator of $\xi_{\rm NL}$, no new measurements were required to estimate $\xi_{\rm Lag}$.  This is in contrast to all previous work, e.g.  Ref.~\cite{LPlaguerre}, where it was necessary to fit each of the $\xi_{\rm NL}$ curves shown in Figure~\ref{fig:xibinned} using modified Laguerre functions, and then insert the best-fit coefficients in the second of equations~(\ref{eq:xiLag}) to estimate $\xi_{\rm Lag}$.  In our case, our estimation of $\xi_{\rm NL}$ is done directly in terms of  modified Laguerres (i.e. the curves in Figure~\ref{fig:xinobins}, rather than Figure~\ref{fig:xibinned}), so determining the best-fit coefficients is not a separate step.  This demonstrates explicitly that, by projecting $\xi_{\rm NL}$ onto the modified Laguerre-functions, it is straightforward to effectively combine the measurement and reconstruction steps into one.


\section{Discussion and conclusions}\label{sec:discuss}
We derived a no-bins estimator of the pair correlation function $\xi$ (Figure~\ref{fig:xinobins}) which accounts for the covariance that arises from Poisson discreteness and the finite volume of the survey (equation~\ref{eq:full}).  The error bands on this estimate are also straightforward to obtain (equations~\ref{eq:covA} and~\ref{eq:covXiNL}).  

Although our formulation is useful for any set of basis functions, some basis functions are more useful than others.  This is generically true for functions which diagonalize one or another of the matrices which appear in equation~(\ref{eq:covA}), and it is particularly true in the context of problems where the measured pair counts have been smeared by a known kernel, and one is interested in deconvolving or reconstructing the pair counts prior to smearing (Appendix~\ref{sec:smearing}).  

For smearing by an isotropic Gaussian kernel, the half-integer generalized Laguerre functions of Ref.\cite{LPlaguerre} are particularly interesting: their use effectively combines the measurement of the non-linearly evolved $\xi_{\rm NL}$ and its Laguerre-reconstructed linear theory shape $\xi_{\rm Lag}$ into a single step (Figure~\ref{fig:xirecon}).  Moreover, in the context of cosmological distance scale estimates, using these functions in our framework shows how the uncertainty bands around $\xi_{\rm NL}$ and $\xi_{\rm Lag}$ are related to one another, and to the survey discreteness and volume, in a fairly transparent and intuitive way (compare equations~\ref{eq:covA} and related discussion).  Another interesting set is the `Bi-Sequential' basis of Ref.\cite{biseq}, or their Gaussian smoothed versions, as these provide a model-agnostic framework for fitting a wide variety of cosmological models.  In practice, reconstruction using either the Laguerre or Bi-Sequential bases requires an estimate of the smearing scale (Figure~\ref{fig:xirecon} assumed this scale was known perfectly), and marginalizing over uncertainties in this scale is slightly cumbersome. Although we did not exploit it here, this marginalization is simpler in the approximation discussed by Ref.\cite{LPhermite}, where $\xi_\text{NL}$ would be projected onto a basis of simple polynomials, and the associated reconstruction would use a set of modified-Hermite functions. 

The linear point scale ($r_{\rm LP}$ of equation~\ref{eq:rLP}) measured in $\xi_{\rm Lag}$ is more like that in linear theory than is $r_{\rm LP}$ in $\xi_{\rm NL}$ (Figure~\ref{fig:rLPvEff}).  Moreover, these $r_{\rm LP}$ scales and their associated errors (equation~\ref{eq:sigLP}) are similar to those obtained from estimating $\xi_{\rm NL}$ in bins, and then performing Laguerre reconstruction as a second separate step (discussion following Figure~\ref{fig:xirecon}).  By combining the two steps into one, our no-bins estimator of $\xi$ has simplified the process of Laguerre reconstruction while preserving the gains in accuracy and precision that reconstruction provides.  (This would be especially true for the polynomial - modified-Hermite combination mentioned above.)

Finally, by exploiting the fact that $\xi$ is a relatively smooth function of scale, our approach expresses the shape of $\xi$ using many fewer free parameters than traditional binned estimators.  E.g., Ref.\cite{bayesLP} suggest that $N\sim 7$ Laguerre functions rather than $N\sim 30$ tophat bins may be sufficient for the next generation of datasets.  The computational load for estimating the elements of the covariance matrices that are required for quantifying the uncertainties on cosmological constraints scales as $N(N+1)/2$, so our method reduces this load dramatically.  Therefore, we expect it to be useful in analysis of the next generation of cosmological distance scale measurements.  Of course, this also motivates generalizing the method here to higher order statistics (e.g., the three-point function), where the potential gains in covariance matrix analyses are much more significant, and to problems where the smearing kernel is not isotropic.  

\begin{acknowledgments} 
  FN, MM and RKS thank the Munich Institute for Astro- and Particle Physics (MIAPP) which is funded by the Deutsche Forschungsgemeinschaft (DFG, German Research Foundation) under Germany's Excellence Strategy – EXC-2094 – 390783311, for its hospitality during the summer of 2019.  MM and RKS thank the ICTP and the IFPU for their hospitality in Trieste during the summer of 2021.  RKS thanks EAIFR, IUCAA and IFPU for their hospitality in summer 2024, spring 2026 and summer 2026 when this work was (finally!) completed.  
  FN acknowledges support from the National Science Foundation Graduate Research Fellowship (NSF GRFP) under Grant No. DGE-1845298, and from the Yale Center for Astronomy and Astrophysics Prize Postdoctoral Fellowship.
  MM is supported by the grants PID2024-158938NBI00 and CNS2024-154286 funded by the Agencia Estatal de Investigación of the Spanish Ministerio de Ciencia, Innovación y Universidades, (MICIU/AEI/10.13039/501100011033) and “ERDF A way of making Europe”, and by the Project SA097P24 funded by the Junta de Castilla y Léon.
  The research of AP is supported by the Associates Scheme of ICTP, Trieste.
\end{acknowledgments}

\bibliography{nobinXi}

\appendix

\section{Gram-Schmidt orthogonalization} \label{sec:gso}
The main text notes that there is some merit to working with orthogonal basis functions.  Any set of basis shapes $f_k(r)$ can be made orthogonal by working with  
\begin{equation}
 g_n(r) = {\rm Det} \begin{vmatrix}
          F_{00} & F_{01} & \ldots & F_{0n} \\
          F_{10} & F_{11} & \ldots & F_{1n} \\
          \vdots & \vdots & \ddots & \vdots \\
          F_{n-1,0} & F_{n-1,1} & \ldots & F_{n-1,n} \\ 
          f_0(r) & f_1(r) & \ldots & f_n(r)
          \end{vmatrix} .
\end{equation}
Typically $f_0(r)=1$ for $[r_{\rm min},r_{\rm max}]$ and is zero outside, so $F_{0j} = v_j$.  Therefore, $g_0(r) = 1$, and its volume is $v_0$ (the same as for $f_0$).  The volumes of all the other $g_j(r)$ vanish (i.e. are equal to zero).  This is most easily seen by noting that the volume of $g_j(r)$ is got by integrating it over $r$.  In effect, this means that the $f_j$ in the matrix expression above should be replaced with $v_j$.  However, since $v_j = F_{0j}$, this makes the top and bottom rows equal; the determinant of a matrix with two identical rows is zero.  

The vanishing of all $v_j$ except $v_0$ significantly simplifies equations~(\ref{eq:simple}) and~(\ref{eq:full}) in the main text.  They become 
\begin{equation}
    a_0 = \frac{DD_0 - v_0}{G_{00}} = \frac{DD_0}{v_0} - 1 \quad {\rm and}\quad 
    a_k = \frac{DD_k}{G_{kk}} ;
\end{equation}
this provides yet another reason why orthogonal bases are attractive.  
Since $DD_0$ is just the fraction of pairs that have separations between $r_{\rm min}$ and $r_{\rm max}$, $a_0$ is the integral of the pair correlation function over the volume between $r_{\rm min}$ and $r_{\rm max}$, divided by the volume between $r_{\rm min}$ and $r_{\rm max}$.  

Finally, note that when $P(k)=0$, then the only contribution to ${\bm E}$ is from the shot-noise term ${\bar n}^{-1}\equiv V_{\rm tot}/N$:
\begin{align}
    E_{ij} &\to \frac{2}{{\bar n}^2V_{\rm tot}} \int_0^\infty \frac{dk\,k^2}{2\pi^2}\,w_i(k) \, w_j(k) \nonumber\\ 
    &= \frac{2}{{\bar n}^2V_{\rm tot}} \int_0^\infty \frac{dk\,k^2}{2\pi^2}\, 
     \frac{4\pi}{V_{\rm tot}} \int dr\, r^2\, f_i(r)\,j_0(kr)\nonumber\\
    & \quad\times\quad \frac{4\pi}{V_{\rm tot}} \int ds\, s^2\, f_j(s)\,j_0(ks)\nonumber\\
    &= \frac{2}{{\bar n}^2V_{\rm tot}}\,\frac{4\pi}{V_{\rm tot}^2}\int dr\, r^2\, f_i(r)\,f_j(r)\,\nonumber\\
    &= \frac{2}{{\bar n}^2V_{\rm tot}^2}\,F_{ij} .
\end{align}
For orthogonal basis functions $g_n$, we set $\bm{F}\to\bm{G}$ which is diagonal, so $\langle\xi(\bm{r})\xi(\bm{s})\rangle \to (2/N^2)\,\sum_j g_j(r)g_j(s)/G_{jj}$.  
Therefore, in the limit in which shot-noise dominates, there may be significant gains to be had by working with orthogonal basis functions.

\section{Basis functions and smearing}\label{sec:smearing}
Suppose that $\xi(s)$ of equation~(2) can be written as 
\begin{equation}
    \xi(\bm{s}) = \int d\bm{r}\, \xi_{\rm L}(\bm{r})\,K(\bm{r},\bm{s})
\end{equation}
with 
\begin{equation}
    \xi_{\rm L}(\bm{r}) = \sum_j a_j\, p_j(\bm{r})
\end{equation}
for some basis functions $p_j(\bm{r})$.  We will refer to $K$ as the `smearing kernel'.  For example, Ref.~\cite{LPlaguerre} set 
\begin{equation}
    K(\bm{r},\bm{s}) =  \frac{{\rm e}^{-(\bm{s}-\bm{r})^2/2\Sigma^2}}{(2\pi\Sigma^2)^{3/2}} 
\end{equation}
and 
\begin{equation}
    p_j(\bm{r}) = \left(\frac{|\bm{r}|-r_{\rm fid}}{\sigma}\right)^j.
\end{equation}
Then, if we use $q_j(\bm{s})$ to denote the smeared value of $p_j$ (i.e. the result of the convolution integral of $p_j$ over $\bm{r}$), we have 
\begin{equation}
    \xi(\bm{s}) = \sum_j a_j\,q_j(\bm{s}).
\end{equation}
Equation~(\ref{eq:dk}) becomes 
\begin{align}
    d_k &= \int \frac{d\bm{s}}{V_{\rm tot}}\, f_k(\bm{s})\,\big[1 + \xi(\bm{s})\big] \nonumber\\
    &= v_k + \sum_j a_j \int \frac{d\bm{s}}{V_{\rm tot}}\, f_k(\bm{s})\,q_j(\bm{s}),
\end{align}
where $v_k$ is given by equation~(\ref{eq:vk}).  
If we now set $f_k = q_k$, then the integral on the final line becomes $F_{jk}$ of equation~(\ref{eq:simple}).  We use this in the main text.  Note, however, that if the kernel is symmetric in $\bm{r}$ and $\bm{s}$, then $f_k = p_k$ would also be a natural choice.  E.g., for the Gaussian smearing kernel, 
\begin{align}
    d_k 
    &= v_k + \sum_j a_j \int \frac{d\bm{s}}{V_{\rm tot}}\, f_k(\bm{s})\,\int d\bm{r}\, p_j(\bm{r})\,
     \frac{{\rm e}^{-(\bm{s}-\bm{r})^2/2\Sigma^2}}{(2\pi\Sigma^2)^{3/2}} , 
\end{align}
showing that $f_k = p_k$ treats the two basis functions equally.  In this case, one would simply modify the definition of $F_{jk}$ in the main text to now equal 
\begin{equation}
    F_{jk} = \int \frac{d\bm{s}}{V_{\rm tot}}\, f_k(\bm{s})\,q_j(\bm{s}).
\end{equation}
The rest of the analysis in the main text is unchanged: The fitted $a_k$ coefficients yield $\xi(r)=\bm{a}^{\rm T}\bm{q}(r)$ and $\xi_{\rm rec}(r)=\bm{a}^{\rm T}\bm{p}(r)$ as the estimated actual and reconstructed correlation functions.

\end{document}